-RESEARCH ARTICLE-

# HERD BEHAVIOR IN CRYPTO ASSET MARKET AND EFFECT OF FINANCIAL INFORMATION ON HERD BEHAVIOR

**Üzeyir AYDIN**
Associate Professor, Dokuz Eylül University, Faculty of Economics and Administrative Sciences, Economics, İzmir, Turkey
**EMAIL**      : uzeyir.aydin@deu.edu.tr
**ORCID ID**   : https://orcid.org/0000-0003-2777-6450

**Büşra AĞAN**
Research Assistant, Eastern MediterraneanUniversity, Faculty of Business and Economics, Economics, Famagusta via Mersin 10, Turkey.
**EMAIL**      : busra.agan@emu.edu.tr
**ORCID ID**   : https://orcid.org/0000-0003-1485-9142

**Ömer AYDIN**[1]
Lecturer Dr., Dokuz Eylül University, Faculty of Economics and Administrative Sciences, İzmir, Turkey
**EMAIL**      : omer.aydin@deu.edu.tr
**ORCID ID**   : https://orcid.org/0000-0002-7137-4881

**-Abstract-**
The initial purpose of the study is to search whether the market exhibits herd behavior or not by examining the crypto asset market in the context of behavioral finance. And the second purpose of the study is to measure whether the financial information stimulates the herd behavior or not. Within this frame, the announcements of Federal Open Market Committee (FOMC), Governing Council of European Central Bank (ECB) and Policy Board of Bank of Japan (BOJ) for interest change, and S&P 500, Nikkei 225, FTSE 100 and GOLD SPOT indices'

---

[1] **Corresponding Author:** omer.aydin@deu.edu.tr

**Citation (APA):** Aydın, Ü., Ağan B. & Aydın Ö., (2020), Herd Behavior in Crypto Asset Market and Effect of Financial Information on Herd Behavior, International Journal of Economics and Finance Studies, 12(2):581-604. Doi: 10.34109/ijefs.202012221

data were used. In the study, the analyses were made over 100 cryptocurrencies with the highest trading volume by the use of 2014:5 - 2019:12 period. For the analysis, the Markov Switching approach as well as loads of empiric models developed by Chang et al. (2000) were used. According to the results obtained, the presence of herd behavior in the crypto asset market was determined in the relevant period. But it was found that interest rate announcements, and stock exchange performances had no effect on the herd behavior.

## 1. INTRODUCTION

Traditional finance theories assume that the individuals behave rationally, and tell about how they should behave for maximizing their assets. Rational person is the one who is able to update her/his expectations in the light of new information on the assumptions of effective markets hypothesis, who has all kinds of clear and sufficient information, who simultaneously accesses the new information arriving the market, and who doesn't repeat the same mistakes within the frame of expected utility theory while making investment selection.

The reason of arise of behavioral finance is based on the facts that the individuals are not always rational, and that they don't act consistent in their decisions, and thus that the theories such as Efficient Markets and Prospect Theory are not valid. In traditional finance, the investors are devoid of the emotions such as self-control and remorse. Thus, rationality is a human behavior assumption lying at the basis of economical research, and directing it. But it is being anticipated in both economics researches and psychology researches that the individuals don't behave rationally in any case, and that the deviations from rationality are systematic and predictable. While the small researches made bring along great results, the rationality of individuals, and its consistency with the analyses made in the decision making process have started to be searched. While the analyses made have unveiled some behavioral tendencies, explanations have been tried to be made within the frame of these tendencies for understanding the financial markets and institutions.

In 1970s, while the cognitive psychologists had started with studies regarding economic decision making, by selecting the maximization of expected utility and Bayesian probability judgment as the baselines, they had theorized the deviations from those baselines by cognitive mechanisms through approaches different from the one suggested by the behavioral economist Simon (Camerer, 1999: 1). Those studies had drawn the attention of many economists and psychologists, and had gave rise to the formation which would then be called as second generation behavioral economists, or new behavioral economists. The leaders of new behavioral economics are A. Tversky and D. Kahneman who were cognitive psychologists. Moreover, other behavioral economists, in the development period of new behavioral economics as from 1970s until 1980s, are known as P.Solvic and R. Thaler. And after 1990s, C. Camerer S. Mullanaithan, G. Lowenstein, D.Laibson, ve M. Rabin have contributed to the development of today's behavioral economics (Eser and Toigonbaeva, 2011:298-299).

Behavioral finance, while explaining the financial decision making behavior, defends an innovative approach in the discipline of economics by expressing that the individuals may behave irrationally, and that their behaviors may differentiate under different conditions and environments through revealing some perceptions and the importance of emotional intelligence. In behavioral finance, analysis is not just being made with mathematical data, and theories are being formed, and it is being progresses by adding psychological, sociological and anthropological factors to the referred theories. Risks and uncertainties, presence of asymmetric information, and psychological reasons such as desire for status and ownership, being popular, liked and respected, and tendency to keep up with the herd are creating differentiations in the behaviors of individuals.

Kahneman and Tversky (1979) had laid the foundations of behavioral finance by revealing the Prospect Theory as the result of some approaches suggesting that the human brain is making decisions as being affected by psychological and sociological factors under risk and uncertainty. Actually, Kahneman and Tversky suggest, rather than the irrationality of individuals, that the individuals' intuitions, perceptions, urge of avoiding from asymmetric information, uncertainty and risk, desire of gaining status and reputation, or fear of losing, and cognitive abilities operate so well that these features make them fail under some contexts, and cause systematic mistakes. Thus, predictable irrational behaviors arise (Sunstein, 2016). In general terms, behavioral finance, by revealing the importance of working interdisciplinary and multidisciplinary with social sciences, specifies that the individuals are limited rational while making their economic decisions, that they might not always be in pursuit of maximization of utility, and that sociological and psychological factors may also play an effective role in decision making behavior (Lewis et al., 2009: 432). According to Simon, universal rationality is not possible. Because the organism's knowledge and abilities will limit the rationality. By the words of Simon, it should be dealt with "limited" rationality models not with "universal" rationality models (Simon,1955:112-113). At this point, it is beneficial to differentiate objective and subjective rationality (Aydın and Ağan, 2018: 278). Objective rationality is based on the assumption of "rational person" being known in traditional economics. Subjective rationality is a concept being used for explaining that everyone is rational in their opinion, but an individual behaving in this manner may seem irrational when observed by an objective eye. For instance, an economic unit, which is thinking that its knowledge and experience is insufficient, and for that reason which is tending to herd behavior by behaving in the direction of keeping up with the majority, is acting in this way as it is thinking that it is rational. But an individual observing with an objective eye may anticipate that it is irrational. In this case, the economic unit is able to be irrational as per an objective perspective while it is rational subjectively. In this sense, especially the crypto asset markets are inclined to critics that the behavioral finance supporters suggest for financial markets.

Crypto assets are known as currency systems that are gradually becoming widespread in the whole world, that cannot be controlled by a government, company or authority, and that are not central, and as currencies that are using the science of cryptology, and that are digital and virtual encrypted on mathematical basis. Cryptology is a science of encryption. They are able to be used from the virtual wallets, in which they are placed by the use of specific ciphers, via the ciphers. Thus, the individuals obtain a real expenditure and gain via these assets. The value of crypto assets is being determined by the perception of it by the users as a means of exchange and commodity under instant supply and demand conditions in the market. The objective of Bitcoin, being the most successful crypto asset until today, is actually in the same direction with fintech companies, and it is decreasing the transaction costs and completely eliminating the need for financial intermediaries. Bitcoin, which is the first distributed model, had been suggested in 2008 by Satoshi Nakamoto as an innovative payment system, and as a new digital currency. It had been made ready for use in an environment when trust against intermediary firms, banks and central banks, and governments had decreased. Such that, Bitcoin expresses the virtual currency being used in the markets via internet, and only one of the assets which is being independent from central authority or intermediary firms, in other words which is called as crypto asset (Ağan and Aydın, 2018: 798). Following Bitcoin, many other practices have also been introduced. These are based on blockchain technology, and sometimes they are the copies of Bitcoin, but they may differ in many aspects. These are being called subcoins which are alternatives of other crypto assets (Bonneau et al., 2015). Ethereum, Ripple,

Litecoin, Dash, Monero, Neo, Nem may be given as examples of popular subcoins. Just like the stocks and options, Bitcoin and subcoins are also able to be exchanged in crypto asset markets. But this exchange doesn't arise from medium of exchange structure of crypto assets, their value, and their functioning as a unit of account. Crypto assets are unable to completely have these features due to their price volatility in the market. Because the volatility of crypto assets is higher compared to other currencies, and they are able to expose the users to short-term risks (Şanlısoy and Çiloğlu, 2019). The low correlation of cryptocurrency with daily foreign exchange rates, with extensively used currencies, and with gold is making the crypto asset ineffectual for portfolio management. Exposure of crypto assets to risks of daily attacks and theft, inability of them to be used in conventional financial transactions, inability to provide transparency of information, lack of legal authority and a strong legal frame behind it are making the crypto asset as a speculative investment tool rather than a currency (Yermack, 2013). The main reason of this assessment is the excessive returns of crypto assets due to the periodical high volatility of crypto asset prices. This state is able to cause the presence of speculative bubbles in crypto asset market. Many studies made are of a character proving the same (Phillips, Shi and Yu, 2013; Kindleberger, Aliber and Wiley, 2005; Cheah and Fry, 2015; Cheung et al., 2015; Corbet et al., 2017; Blau, 2017; Balcilar et al., 2017; Urquhart, 2018; Ceylan et al., 2018). A part of these studies had found out that announcements for loosening the monetary policies in USA, EU, UK and Japan, and that monetary policy decisions based on interest rates are causing volatility on the prices of Bitcoin. High demand arising in periods of price bubbles is able to increase the prices more, or able to bring down more the rapid price decreases. Within this frame, the main motivation behind increase of demand is causing of herd behavior by the speculative buyers through development of action model "with the fear of missing the opportunity / losing" (Corbet et al., 2017: 62).

However, according to classic theories of economics, as the economic agents make decisions by using all the available information, the idea regarding the possibility of arise of an irrational market directed by the herds is completely being eliminated. An opposite opinion is that financial investment arises from herd behavior weakening the connection in between knowledge and results of market (Devenow and Welch 1996; Scharfstein and Stein 1990). For this reason, behavioral finance, by examining the beliefs and decisions of investors in the real world, tries to reveal the results of market in the presence of a large irrational investor group. The individuals, dealing with crypto asset units, are generally being fed from two sources while making their investment decisions: news and social media. Today, many forums, where issues relevant to referred asset markets are being discussed in a wide array, have been formed. The investors are sharing and following-up the recent subjects such as the recent news, unexpected increases or decreases in crypto asset prices, innovations on Blockchain platform on these forums. Thus, as it will also be understood by the evaluation of literature, through the formation of ideas by the community within the forum, different investment strategies such as discovering a new subcoin or defining "smart" price models are being revealed, and herd behavior is able to be actualized.

In this context, in this study it is being focused on the fact that the behaviors of investors are not in conformity with a rational criterion, and that these behaviors are activating a collective decision making process by forming a coordination mechanism in the formation and determination of crypto asset prices. Based on this focus point, the main purpose of the study is to search whether the crypto asset market shows herd behavior or not under hypothesis that investors of crypto asset have limited sources and weak prior knowledge for processing the information. And the second purpose of the study is to measure whether the financial information stimulates the herd behavior or not. Within this frame, the study was constructed as follows: First, literature analyzing the herd behavior was included. And then, the data set used for the study was explained. In the third section, methodology relevant to herd behavior, and empiric results were indicated. And the study was finalized by the conclusion section where the main results and policy suggestions are present.

## 2.   LITERATURE REVIEW

Within the frame of behavioral finance, herd behavior is a decision making approach which is being characterized by mimicking the actions of others (Hirshleifer and Hong Teoh 2003).  The herd behavior, which is being characterized as a fashion or as a fancy, is the tendency to keep up with the majority. It is individuals' actions according to the decision of the group they are involved in by ignoring their own decisions.  In cases when the individuals are required to make a decision, it is the individuals' rapport with the preferences of the group through consideration that decisions made by the majority is healthier than their personal decisions. Herd behavior is one of the most effective behavioral tendencies in the decision making process of the individuals for investment (Hotar, 2020: 87). For this reason, studies of financial economists and financial investors for understanding the herd behavior have recently increased at a great extent. Christie and Huang (1995); Chang et al. (2000); Hwang and Salmon (2004), Gleason et al. (2004), Demirer et al. (2010), Chiang and Zheng (2010), Cheah and Fry (2015), Urquhart (2017), Katsiampa (2017), Poyser (2018), Bouri et al. (2018),  Ajaz and Kumar (2018), Da Gama Silva et al. (2019),  Hotar (2020) had proved the presence of herd behavior in their studies by which they have searched the presence of herd behavior.

Bourie et al. (2018) had determined the presence of a herd behavior which is changing in time in the crypto asset market by the use of Cross-Sectional Absolute Deviation (CSAD) criterion of Chang et al. (2000), and of the rolling window method for fourteen cryptocurrencies. Poyser (2018) had determined that the investors are following the community in periods of uncertainty in the market by the use of CSAD criterion of Chang et al. (2000), and of Markow Switching approach for 100 cryptocurrencies. Ajaz and Kumar (2018) had determined the presence of herd behavior in crypto asset market by the use of CSAD criterion of Chang et al. (2000) for six cryptocurrencies, and had expressed that it is indicating excessive reaction. Vidal-Thomas et al. (2018) had determined that herd behavior is in subject in descending markets as per the CSAD criterion by the use of Cross-Sectional Standard Deviation (CSSD) suggested by Christie and Huang (1995), and of CSAD criterion suggested by Chang et al. (2000) for sixty five crypto assets, and they had asserted that this state is increasing the ineffectiveness of the markets and risk of making investment, and that subcoins are mimicking the large cryptocurrency investments. King and Koutmos (2018), in their research performed for nine crypto assets, had determined that the herd behavior is in subject, and that the delayed values are causing the arise of herd behavior. Da Gama Silva et al. (2019) had searched the herd behavior by the use of CSAD, CSSD, and the criteria used by Hwang and Salmon (2004). Consequently, they had determined the herd behavior. Hotar (2020) had searched the herd behavior tendency on the market of twenty-two crypto assets, having highest trading volume, by the use of CSAD criterion and of Markow Switching approach. In the research, they had proved the presence of herd effect in periods of ascending market and high volatility. Lieure (2018), who had used a different method by the herd behavior model based on knowledge, had searched the presence of herd behavior by means of Google Trends. He had found a significant positive relationship in between herd behavior and event uncertainty. Asteriou (2018) had examined the herd behavior by structure breakdown tests, and had determined the presence of herd behavior. But he couldn't prove the presence of herd behavior in days of negative returns.

Briefly, it is being observed that crypto assets exhibit excessive return and volatility from time to time without being based on information. Inexperienced investors, relying on news and information whose accuracy had not been proven, are actualizing crypto asset transactions without completely measuring the risks. Actually, the investors are being affected from the transactions of others as independent from their own analyses. It has been specified in literature that this state may cause intensifying potential herd behavior due to uncertainties and market conditions.

## 3. DATA SET

According to the data of CoinMarketCap, there were more than 5 thousand crypto assets in the market until March 2020, and their number is continuously increasing. In this study, 100 crypto assets, with the highest trading volume and constituting about 90% of all the crypto assets, were included in the analysis. The data set is covering the period of 2014:5-2019:12. The closing prices of crypto assets were obtained from the internet address of https://coinmarketcap.com. For the analysis, the Markov Switching approach as well as Cross-Sectional Absolute Deviation (CSAD) developed by Chang et al. (2000) were used. In order to test the presence of herd behavior, variables of Cross-Sectional Absolute Deviation (CSAD) coefficient, and absolute value of the difference of return of cryptocurrency ($R_{it}$) and equal weighted market return ($R_{mt}$) were used. Moreover, depending on the other purpose of the study, interest rates news, index returns and golden returns were used for the determination of whether the macroeconomic and financial data activate the herd behavior or not. The interest rate announcements, obtained from the official internet pages of Federal Open Market Committee (FOMC) in USA – for the interest rate news of Federal Reserve System (FED)-, of The Governing Council of The European Central Bank (ECB) –for the interest rate news of European Central Bank-, of The Policy Board of the Bank of Japan (BOJ) in Japan

–for the interest rate news of the Central Bank of Japan-, were used. And S&P 500, Nikkei 225, FTSE 100 and GOLD SPOT indices data was obtained from the address of investing.com. All data was used in the analyses as being cleared of seasonality. The information on variables used in the study, and unit root tests used for the determination of stagnation levels are given in Table 1. As it can be observed from Table 1, according to Augmented Dickey-Fuller (ADF) and Philips-Perron (PP) unit root tests, the series are stagnant at the level of I(0).

**Table 1:** Data Set, and Analysis Results of Unit Root Tests

| Variables | Remarks | ADF Test | | PP Test | |
|---|---|---|---|---|---|
| | | Model with Constant | Model with Constant + Trend | Model with Constant | Model with Constant + Trend |
| Coefficient of Cross-Sectional Absolute Deviation (CSAD) | $CSAD_t = \frac{1}{N}\sum_{i=1}^{N} |R_{i,t} - R_{m,t}|$ | -8.120*** (0.00) | -9.982*** (0.00) | -8.986*** (0.00) | -9.560*** (0.00) |
| Return of equal weighted market portfolio ($R_{m,t}$) | $R_{mt} = \frac{\sum_{i=1}^{N} e_{i,t}}{N}$ | -8.561*** (0.00) | -9.459*** (0.00) | -8.239*** (0.00) | -8.451*** (0.00) |
| Absolute value of the difference of return of equal weighted market portfolio ($R^2_{m,t}$) | Absolute value of the difference of return of equal weighted market portfolio was calculated | -9.051*** (0.00) | -9.257*** (0.00) | -9.567*** (0.00) | -9.346*** (0.00) |
| Return of S&P 500 Index (S&P 500) | % difference values were used | -10.735*** (0.00) | -10,659*** (0.00) | -11.198*** (0.00) | -11.280*** (0.00) |
| Return of Nikkei 225 Index (Nikkei 225) | % difference values were used | -8.977*** (0.00) | -6.131*** (0.00) | -8.532*** (0.00) | -8.898*** (0.00) |
| Return of FTSE 100 Index | % difference values were used | -12.138*** (0.00) | -10.131*** (0.00) | -12.705*** (0.00) | -12.890*** (0.00) |
| Return of GOLD SPOT (GOLD SPT) | % difference values were used | -6.033*** (0.00) | -6.295*** (0.00) | -6.631*** (0.00) | -6.724*** (0.00) |
| Interest rate announcements of FOMC, ECB and BOJ ($XR^2_{m,t}$) | In case making interest decision (interest rate increase, decrease, or not making any change) 1; if not 0. | | | | |

Note: The values in the tests within parenthesis represent the p value. *: Significant at the level of 1%.

## 4. METHOD AND FINDINGS

In literature, even if it is being specified that direct observation of the actions of the investors is the best approach for testing the herd behavior, direct observation is nearly impossible due to the confidentiality being present in the crypto asset market. For this reason, as specified in the section of literature, a few methods were developed for testing the herd behavior. Lakonishok et al. (1992) had made the first contribution to the methodology by the LSV measurement model. Wermers (1995) had used the portfolio change criterion of correlated transactions covering the intensity of the purchase and sales transactions of the investors. And Christie and Huang (1995) had applied a new methodology based on the deviation of the returns of share certificates. In the study, they had tested the herd behavior by examining the cross-sectional deviations of share certificate returns in USA stock exchange markets according to the average of market. According to Christie and Huang (1995), as the distribution measures the proximity of returns to the average, when the personal returns follow-up the leadership of portfolio return, then it reveals the presence of herd behavior. And Chang et al. (2000), based on the results of the model of Christie and Huang (1995), had tested the herd behavior by developing a new model covering the nonlinear relationship in between the share certificate return deviation and market return. In that model, it had been intended to determine any possible nonlinear relationship in between share certificate return distributions and market return.

As is seen, different proxies had been developed in literature for determining the herd behavior. In this study, the methodology of Chang et al. (2000), being a development of the original methodology presented by Christie and Huang (1995), was used. Christie and Huang (1995) had suggested the use of Cross-Sectional Standard Deviation (CSSD) for defining the herd behavior in financial markets. CSSD had been developed in order to capture the differences in the behaviors of investors upon excessive increases or decreases in the markets. However, CSSD has two main disadvantages. First of all, it is very sensitive against inconsistent values. As a second point is that the thing being deemed as "excessive" is completely arbitrary. For this reason, Chang et al. (2000) had modified the model of Christie and Huang, and had developed the Cross-Sectional Absolute Deviation (CSAD) model which is being formulated as follows.

$$CSAD_t = \frac{1}{N}\sum_{i=1}^{N}|R_{i,t} - R_{m,t}| \qquad (1)$$

$R_{i,t}$: Daily return for each crypto asset of N number

$R_{m,t}$: Absolute value of equal weighted market return difference

CSAD is a distribution measure considering the absolute difference in between the personal return and average market returns. The researchers had constructed the model on the following information: "If the participants of the market ignore their own priorities by following-up the behaviors of the market in periods of great price movements, then the linear and increasing relationship in between the financial asset return distribution and market return will not be valid anymore. Instead, the relationship will be in a nonlinear manner." The model constructed on this argument had been empirically tested by some researchers such as Arjoon and Shekhar (2017), Chiang and Zheng (2010), Demirer et al.(2015), Balcılar et al. (2013), Poyser (2018). Following the referred studies, the equation formed for method based on cross-sectional deviations of crypto asset returns is as follows:

$$CSAD_t = \Upsilon_0 + \Upsilon_1|R_{m,t}| + \Upsilon_2 R_{m,t}^2 + \varepsilon_t \qquad (2)$$

In the estimation, as the determination of any possible nonlinear relationship in between the crypto asset return distributions and market return was intended, in Equation 2 if the $(\Upsilon_2)$ coefficient of nonlinear term $(R_{m,t}^2)$ is negative and statistically significant, then it will indicate the presence of herd behavior, and if it is positive, then it will indicate the lack of presence of herd behavior.

On the other hand, herd behavior is able to change depending on time. In the study, Markov Switching (MS) approach was used for determining whether there exists specific periods in which the herd behavior arise or not, and for determining the regimes in which the herd is present. MS regression provides extremely useful estimators in high frequency nonlinear data (Aydın and Kara, 2014:36).

Below, models for obtaining the estimators are provided. The first model is the standard static herd model which is widespread in literature. The second model is the Markov Switching model indicating the herd behavior on multiple regimes. Markovian herd model may be indicated as follows:

$$CSAD_{t,1} = \Upsilon_{0,1} + \Upsilon_{1,1}|R_{m,t}| + \Upsilon_{2,1}R_{m,t}^2 + \Upsilon_{3,1}Vol_t^{CSAD} + \Upsilon_{4,1}Vol_t^{R_{m,t}} + \Upsilon_{4+k,1}CSAD_{t-k} + \varepsilon_{t,1}$$

$$CSAD_{t,2} = \Upsilon_{0,2} + \Upsilon_{1,2}|R_{m,t}| + \Upsilon_{2,2}R_{m,t}^2 + \Upsilon_{3,2}Vol_t^{CSAD} + \Upsilon_{4,2}Vol_t^{R_{m,t}} + \Upsilon_{4+k,2}CSAD_{t-k} + \varepsilon_{t,2}$$

$$CSAD_{t,s} = \Upsilon_{0,s} + \Upsilon_{1,s}|R_{m,t}| + \Upsilon_{2,s}R_{m,t}^2 + \Upsilon_{3,s}Vol_t^{CSAD} + \Upsilon_{4,s}Vol_t^{R_{m,t}} + \Upsilon_{4+k,s}CSAD_{t-k} + \varepsilon_{t,s}$$

$\varepsilon_{t,s} = N(0, \sigma_s^2)$. Transition probabilities $p_{i,s}$ are defined as the probability switching from regime $s = 1,2,3$ to regime $i = 1,2,3$ at time t.

Due to the differences in the dynamic structures of the examined variables, and data dependant structure of CSAD model, number of regimes and definitions of regimes may differ for each series (Aydın and Kara, 2014:39). In the estimated models, number of regimes is 4 for each series as being adhered to the study of Chang et al. (2000). In the model, Regime 1 indicates the period in which the volatility is the highest; Regime 2 indicates the period in which the volatility is low; Regime 3 indicates the period of best income reached at high volatility; and Regime 4 indicates the period of highest loss at high volatility.

Table 2 reports the estimations for the static model and four regime amendment model according to the aforementioned specification. In the estimated equation, there is no problem of heteroscedasticity and autocorrelation. When all the diagnostic tests in Table 2 are considered, it is being observed that the model is suitable and unproblematic. In addition, the $(Y_2)$ coefficient of the term $(R_{m,t}^2)$ is negative and statically significant in all the models except Regime 2. According to this, herd behavior is in subject in crypto asset market. When the coefficients are considered, it is being observed that the herd behavior is stronger in the period in which the volatility is the highest (Regime 1), and in the period of highest loss at high volatility (Regime 4) (-2.76 and -1.07 respectively), and the coefficient of $(R_{m,t}^2)$ being 0.78 in the Regime 2 which is the period of low volatility indicates lack of arise of herd behavior. These results provide proofs regarding that investors exhibit similar behaviors in periods in which the market is volatile and stressed. The findings obtained are supporting the findings of the literature.

**Table 2:** Results of Regression for the Determination of Herd Behavior

| Term | OLS | Regime | | | |
|---|---|---|---|---|---|
| | | 1 | 2 | 3 | 4 |
| **Constant** | -0.041*** | -0.072** | -0.029*** | -0.032*** | -0.079** |
| | (0.000) | (0.040) | (0.000) | (0.000) | (0.040) |
| **CSAD$_{t-1}$** | -0.745** | -0.534* | -0.978*** | -0.690** | -0.934** |
| | (0.030) | (0.080) | (0.000) | (0.020) | (0.030) |
| **CSAD$_{t-2}$** | -1.023* | -0.975** | -0.780*** | -0.652** | -0.912** |
| | (0.090) | (0.030) | (0.000) | (0.020) | (0.040) |
| **CSAD$_{t-3}$** | -0.675** | -0.908** | 0.785* | -0.968*** | -0.765* |
| | (0.040) | (0.020) | (0.060) | (0.000) | (0.080) |
| $R_{m,t}^2$ | **-0.673*** | **-2.761*** | **0.786**** | **-0.832*** | **-1.073*** |
| | **(0.090)** | **(0.060)** | **(0.050)** | **(0.090)** | **(0.070)** |
| $|R_{m,t}|$ | -0.235** | -2.212** | -0.321 | -0.564** | -0.319** |
| | (0.030) | (0.040) | (0.150) | (0.060) | (0.080) |
| $Vol_t^{CSAD}$ | -0.443 | 1.872 | 1.457 | 0.237 | 1.037 |
| | (0.240) | (0.180) | (0.320) | (0.160) | (0.140) |
| $Vol_t^{R_{m,t}}$ | -0.385 | -0.712 | 0.311 | -0.537 | -0.749 |
| | (0.100) | (0.120) | (0.340) | (0.140) | (0.150) |
| R² | 0.310 | 0.628 | 0.487 | 0.532 | 0.790 |

Note: The values within parenthesis represent the p value. *, **, *** indicate significance at the levels of 10%, 5% and 1% respectively.

After obtaining proofs regarding herd behavior in crypto asset market, the presence of fake herd behavior or intentional herd behavior directed by financial information / news was questioned in the following part of the study. It was searched whether the herd behavior exhibited by the investors transacting at crypto asset markets is being affected from announcements regarding interest rates and from stock

595

exchange performances (index, and gold returns). For this, the following Equation 3 was formed based on the approach of Bikhchandani and Sharma (2000):

$$CSAD_t = \Upsilon_0 + \Upsilon_1|R_{m,t}| + \Upsilon_2 R_{m,t}^2 + \Upsilon_3 XR_{m,t}^2 + \varepsilon_t \qquad (3)$$

**Table 3:** Results of Regression for Interest Rate Announcements

| \multicolumn{7}{c}{$CSAD_t = \Upsilon_0 + \Upsilon_1\|R_{m,t}\| + \Upsilon_2 R_{m,t}^2 + \Upsilon_3 XR_{m,t}^2 + \varepsilon_t$} |
|---|---|---|---|---|---|---|
| Variables | FOMC (+) | ECM (-) | BOJ (-) | FOMC (No change) | ECM (No change) | BOJ (No change) |
| C ($\Upsilon_0$) | 0.042*** (0.00) | 0.048*** (0.00) | 0.053*** (0.00) | 0.059*** (0.00) | 0.078*** (0.00) | 0.090*** (0.00) |
| $R_{m,t}$ ($\Upsilon_1$) | 1.567** (0.02) | 1.675*** (0.01) | 1.876* (0.09) | 1.489*** (0.00) | 1.341*** (0.00) | 1.540** (0.04) |
| $R^2_{m,t}$ ($\Upsilon_2$) | -2.980** (0.03) | -2.164* (0.06) | -1.594* (0.07) | -2.472** (0.03) | -2.279*** (0.01) | 1.702*** (0.00) |
| $XR^2_{m,t}$ ($\Upsilon_3$) | **0.078** **(0.67)** | **0.096** **(0.80)** | **-0.005** **(0.95)** | **0.082** **(0.69)** | **-0.152** **(0.86)** | **0.094** **(0.34)** |
| Corrected $R^2$ | 0.272 | 0.243 | 0.267 | 0.252 | 0.214 | 0.248 |
| F statistics | 8.842*** (0.00) | 8.124*** (0.00) | 8.769*** (0.00) | 8.455*** (0.00) | 8.990*** (0.00) | 8.132*** (0.00) |
| Breusch-Godfrey Autocorrelation LM Test | 1.147 (0.89) | 1.524 (0.56) | 1.649 (0.78) | 1.495 (0.45) | 1.462 (0.18) | 1.972 (0.78) |
| Heteroscedastic ARCH test | 0.642 (0.54) | 0.560 (0.82) | 0.319 (0.68) | 0.431 (0.86) | 0.852 (0.72) | 0.982 (0.48) |

Note: The values within parenthesis represent the p value. *,**,*** indicate significance at the levels of 10%, 5% and 1% respectively.

In the equation, the variable $XR_{m,t}^2$ represents the financial information (interest rate announcements, index, and gold returns). If the referred variable in the regression to be estimated is causing herd behavior, then the coefficient of the variable ($\Upsilon_3$) should be negative and statistically significant. The results regarding the estimation of equation formed by the use of dummy variables, representing the decisions regarding the interest rate announcements of central banks that can make policy design at global level, are shown in Table 3. In the period addressed in the study, while FOMC had made an announcement for interest rate increase, ECB and BOJ had made announcements for interest rate decrease.

As may be seen in Table 3, p values of LM and ARCH tests are bigger than 0.10. According to this, there is no autocorrelation and heteroscedasticity problem in the model, and the model is significant as a whole. And the $(\Upsilon_3)$ coefficient of the variable of $XR_{m,t}^2$ is statistically insignificant. According to this, the finding obtained indicates that the interest rate announcements don't activate the herd behavior in crypto asset market.

In the final section of the study, the effect of stock exchange performance on the herd behavior in crypto asset markets was analyzed. In the direction of this analysis, variable $XR_{m,t}^2$ in Equation 3 represents the index returns. In this study, S&P500, Nikkei 225, FTSE 100 indices, and returns of GOLD SPT were used. Results of analysis are shown in Table 4.

**Table 4:** Results of regression for Index Returns

| | $CSAD_t = \Upsilon_0 + \Upsilon_1|R_{m,t}| + \Upsilon_2 R_{m,t}^2 + \Upsilon_3 XR_{m,t}^2 + \varepsilon_t$ | | | |
|---|---|---|---|---|
| **Variables** | **S&P 500** | **Nikkei 225** | **FTSE 100** | **GOLD SPT** |
| C | 0.045*** | 0.042*** | 0.039*** | 0.042*** |
| $(\Upsilon_0)$ | (0.00) | (0.00) | (0.00) | (0.00) |
| $R_{m,t}$ | 1.103*** | 1.213*** | 1.342*** | 1.505*** |
| $(\Upsilon_1)$ | (0.00) | (0.00) | (0.00) | (0.00) |
| $R^2_{m,t}$ | -4.129 ** | -4.023** | -4.783* | -3.998* |
| $(\Upsilon_2)$ | (0.04) | (0.05) | (0.08) | (0.09) |
| **$XR_{m,t}^2$** | **0.170*** | **0.067** | **-0.087** | **0.052** |
| **$(\Upsilon_3)$** | **(0.10)** | **(0.89)** | **(0.64)** | **(0.72)** |
| Corrected $R^2$ | 0.243 | 0.212 | 0.245 | 0.236 |
| F statistics | 8.282*** | 8.926*** | 8.896*** | 8.480*** |
| | (0.00) | (0.00) | (0.00) | (0.00) |
| Breusch-Godfrey Autocorrelation LM Test | 1.456 (0.43) | 1.564 (0.21) | 1.764 (0.43) | 1.236 (0.27) |
| Heteroscedastic ARCH test | 0.214 (0.64) | 0.331 (0.64) | 0.302 (0.85) | 0.442 (0.79) |

Note: The values within parenthesis represent the p value. *, **, *** indicate significance at the levels of 10%, 5% and 1% respectively.

The model is significant as a whole, and there is no heteroscedasticity and autocorrelation problem. In the model, the coefficient $(\Upsilon_3)$ is either positive, or statistically insignificant. This finding obtained indicates that the index returns don't activate the herd behavior in crypto asset market. In this case, when the results of both analyses are evaluated, it is supporting the argument that herd behavior observed in crypto asset market is not being directed by financial information / news, and that it is an intentional herd behavior arising from exhibition of similar behaviors by the investors. The findings obtained are tallying with the findings of the literature.

## 5. CONCLUSION

In the study, the presence of herd behavior in crypto asset market in the period of 2014:5-2019:12 was analyzed by OLS through the use of CSAD criterion developed by Chang et al. (2000). Moreover, Markov Switching (MS) approach was used for determining the periods in which the herd behavior arises, and for determining the regimes in which the herd is present. As the result of the analysis, it was determined that the herd behavior is present in crypto asset market, and that the herd behavior is stronger during regimes of high volatility and of highest loss at high volatility. And the herd behavior doesn't arise in low volatility regime. The findings obtained are supporting the findings of literature as findings of Hotar (2020) and Poyser (2018) being in the first place.

In the second section of the study, it was examined whether herd behavior exhibited by the investors in crypto asset market is being activated by the interest rate announcements of central banks having the power of affecting the global economy, and by the index returns having the character of being the indicator of global financial performance. By the findings obtained, it was observed that the interest rate announcements and index returns don't affect the herd behavior. This result indicates that the individuals in crypto asset markets don't consider the financial information such as well-attested news, information, basic or technical analysis in cases when they are required to make a decision, but tend to keep up with the majority without completely measuring the risks. In other words, the investors exhibit intentional herd behavior instead of investment behavior directed by financial data as being affected from the transactions of others as independent from their own analyses. The reasons of this may be the following;

- development of a behavioral model by the speculative buyers "with the fear of missing the opportunity/losing",

- following-up of news, shares, comments and ideas on social media and virtually formed forums,

- lack of dependency of crypto asset market on any central authority and on a strong legal frame,

- confidence feelings of younger and inexperienced investors, not having a suitable risk management strategy in crypto asset market, by following-up the winners.

Thus, as long as these reasons are present, herd behavior, which is a strong behavioral tendency, will continue to affect the market. It is required for the policy designers to take the measures which will be able to prevent the possible aggrievement of investors involved in herd behavior by considering the previous market bubbles.